\documentclass[12pt]{iopart}
\usepackage{ae,graphicx}

\input{epsf}

\begin{document}

\title[Dynamical behavior of the Niedermayer algorithm]{Numerical simulation study of the dynamical behavior of the
Niedermayer algorithm}

\author{D. Girardi, N. S. Branco} 
\ead{nsbranco@fisica.ufsc.br}
\address{Departamento de F\' {\i}sica,
Universidade Federal de Santa Catarina, 
88040-900, Florian\'opolis, SC, Brazil} 

\date{\today}

\begin{abstract}

         We calculate the dynamic critical exponent for the 
Niedermayer algorithm applied to the two-dimensional Ising and $XY$ models, 
for various values of the free parameter $E_0$.
For $E_0=-1$ we regain the Metropolis algorithm and for $E_0=1$
we regain the Wolff algorithm. For $-1<E_0<1$, we show that the mean size
of the clusters of (possibly) turned spins initially grows with the linear size of the lattice,
$L$, but eventually saturates at a given lattice size $\widetilde{L}$, which
depends on $E_0$. For $L>\widetilde{L}$, the Niedermayer algorithm is equivalent to
the Metropolis one, i.e, they have the same dynamic exponent.
For $E_0>1$, the autocorrelation time is always greater than for $E_0=1$ (Wolff) and, more
important, it also grows faster than a power of $L$. Therefore,
we show that the best choice of cluster algorithm is the Wolff one,
when compared to the Nierdermayer generalization. We also obtain the dynamic
behavior of the Wolff algorithm: although not conclusive, we propose a
scaling law for the dependence of the autocorrelation time on $L$.

%\noindent     

\end{abstract}

\noindent{ Numerical methods; dynamic behavior; Cluster algorithms; XY model}

\maketitle

%\newpage

\section{Introduction} \label{sec:introduction}

	Numerical simulations have been widely used in the study of physical systems,
specially in the last decades. The field of statistical mechanics, among others,
has benefited a great deal from the use of this technique. In particular,
Monte Carlo methods allowed for a precise determination of thermodynamic parameters
in a variety of models, both classical and quantum. Excellent reviews on these
methods can be found in Refs. \cite{landau} and \cite{newman}.

	In recent years, this field has seen a fast development of new algorithms,
which aim to make the simulation more efficient, both in time and in memory, as
well as broadening its application to more complex systems.
As examples of these developments, we can recall: the calculation of
the density of states through flat histograms, which allows obtaining information
at any temperature from one single simulation, independent of temperature \cite{wang};
the use of bitwise operations and storage, which increases by a great deal
the speed of the update process and saves memory (with the drawback that this
procedure can be used only with specific models) \cite{oliveira}; and the
introduction of cluster algorithms, which updates collections of spins, decreasing
the autocorrelation time and almost eliminating critical slowing down \cite{landau,newman,
swendsen, wolff}.

	In this work we will focus on this last issue. In fact, critical slowing
down is a serious drawback, which makes simulation of systems at, or near, 
critical points very inefficient. This phenomenon is measured through
the scaling of the autocorrelation time, $\tau$, with the linear size of the lattice, $L$,
assumed to be in the form
$\tau \sim L^z$, for points at the critical region. The popular Metropolis algorithm, 
for example, when applied to the Ising model in two dimensions,
presents $z \sim 2.17$ \cite{nightingale}. Algorithms which update clusters of spins 
(the so-called cluster algorithms) have a
much lower value of $z$: this is the case for the Swendsen-Wang \cite{swendsen}
and Wolff \cite{wolff} algorithms, for which $z$ is approximately zero 
for the two-dimensional Ising model \cite{baillie,coddington}.

	An alternative (and generalization) to these last two cluster algorithms, the
Niedermayer algorithm, was introduced some time ago \cite{niedermayer} but,
to the best of our knowledge, has never had his dynamic behavior
studied in detail. In this work, we calculate the dynamic exponent for
this algorithm, applied to the Ising and $XY$ models, for some
values of the free parameter $E_0$ (see below), in
order to determine the best choice of this parameter. 

	This work is organized as follows: in the next section we present
the Niedermayer algorithm and relate it to Metropolis' and Wolff's. In Section
\ref{sec:timeandexponent} we review some features connected to the autocorrelation
time and the dynamic exponent $z$, in Section \ref{sec:results} we present
and discuss our results, and in the last section we summarize the results.

\section{The Niedermayer algorithm} \label{sec:algorithm}

	The Niedermayer algorithm was introduced some time ago and is an option to
Wolff or Swendsen-Wang cluster algorithms. 
The idea is to build clusters of spins and accept their updating
as a single entity,  hopefully in a more efficient
way, when compared to these last two algorithms. In this work,
we have chosen to build the clusters according to the Wolff criterion (they
can be constructed according to the Swendsen-Wang rule but
the results will not differ qualitatively in two
dimensions and in higher dimensions Wolff algorithm in superior
to Swendsen-Wang's). It works as follows, for
the Ising model (the generalization of this algorithm for the $XY$ model
is presented in the Appendix): a spin in the lattice is randomly chosen, being the
first spin of the cluster. This spin is called the \textit{seed}. 
First-neighbours of this spin may be
considered part of the cluster, with a probability
	\begin{equation}
	P_{add} (E_{ij}) = \left\{ \begin{array}{ll}
	                     1-e^{K \left( E_{ij}-E_0 \right) },  & \mbox{if} \;\; E_{ij}<E_0, \\
                             0,                                      & \mbox{otherwise},
	                  \end{array}
                 \right.  \label{eq:Padd}
	\end{equation}
where $K = J/kT$, $T$ is the temperature, $J$ is the exchange constant,
and $E_{ij}$ is the energy between
nearest-neighbour spins in unities of $J$ (i.e, $E_{ij} =- s_i s_j$;
$s_i, s_j=\pm 1$). The free
parameter $E_0$ controls the size of the clusters and the
acceptance ratio of their updating, as seen below. First-neighbours of added spins may be
added to the cluster, according to the probability given above. 
Each spin has more than one chance to be part of the cluster, since it
may have more than one first-neighbour in it. When no
more spins can be added, all spins in the cluster are flipped
with an acceptance ratio $A$. Assuming that, at the frontier of the cluster
there are $m$ bonds linking parallel spins and $n$ bonds linking anti-parallel
spins, $A$ satisfies:
\begin{equation}
   \frac{A(a\rightarrow b)}{A(b\rightarrow a)} = \left[ e^{2 K}
         \left(  \frac{1-P_{add}(-J)}{1-P_{add}(J)} \right) \right]^{n-m}, \label{eq:acceptanceratio}
\end{equation}
where $a\rightarrow b$ represents the possible updating process, 
from the ``old'' ($a$) to the ``new'' ($b$) state, which differ from the
flipping of all spins in the cluster, and $b\rightarrow a$ represents the
opposite move. This expression ensures that detailed balance is satisfied \cite{newman}.

	Now we must consider three cases:
\begin{itemize}
\item [(i)] for $-1 \leq E_0 < 1$, only spins in the same state as the seed
may be added to the cluster, with probability $P_{add} = 1 - e^{-K(1+E_0)}$.
The acceptance ratio (Eq. \ref{eq:acceptanceratio}) cannot be chosen to be one
always and is given by $A = e^{-K(1-E_0)(m-n)}$, if $n<m$ (i.e, if the energy
increases when the spins in the cluster are flipped), or by $A=1$, if
$n>m$ (i.e, if the energy
decreases when the spins in the cluster are flipped). If $E_0=-1$, we obtain 
the Metropolis algorithm, since only one-spin clusters are  possible and the
acceptance ratio is $A = e^{-K \Delta E}$ for positive $\Delta E$ and $1$
otherwise, where $\Delta E = 2 (m-n)$ is the difference in energy when the spin is
flipped, in units of $J$;
\item [(ii)] for $E_0 = 1$, again only spins in the same state can take part of
the cluster, with probability $P_{add} = 1 - e^{-2 K}$. Now, the acceptance
ratio can be chosen to be $1$, i.e, the cluster of parallel spins is always
flipped. This is the celebrated Wolff algorithm;
\item [(iii)] for $E_0 > 1$, spins anti-parallel to the seed
may be part of the cluster,
with probability $P_{add} = 1 - e^{K(1-E_0)}$, while spins in the same
state of the seed have a probability $P_{add} = 1 - e^{-K(1+E_0)}$ of
being added to the cluster. The acceptance ratio is again always $1$. Note
that for $E_0 \gg 1$ nearly all spins will be in the cluster and
the algorithm will be clearly inefficient (in fact, it will not be ergodic for
$E_0 \rightarrow \infty$). Therefore, we expect that, if the optimal choice of
$E_0$ is greater than $1$, it will not be much greater than this value.
\end{itemize}

	Our goal here is to do a systematic study of the Niedermayer algorithm, in order to
establish the optimal value for $E_0$, at least  for the two models addressed
in this text.

\section{Autocorrelation time and dynamic exponent} \label{sec:timeandexponent}

	One possible way to access the dynamic behavior of a numerical
algorithm is to measure the autocorrelation time, $\tau$, of some convenient physical quantity,
which is obtained from the dependence of the
autocorrelation function, $\rho(t)$, on the time $t$. Here, time
is measured in Monte Carlo steps ($MCS$); one $MCS$ is defined as
the attempt to flip $N$ spins, where $N$ is the number of spins in
the (finite) lattice being simulated (in our case, $N=L^2$, where
$L$ is the linear size of the lattice). In fact, a rescaling of the time is necessary,
when dealing with cluster algorithms \cite{newman} and comparing
the results for different values of $E_0$. 
The relation between ``time'' in $MCS$, $t_{MCS}$, and the ``time'' taken to build and possibly flip a cluster, $t$, is
\begin{equation}
   t_{MCS} = t \frac{<n>}{N},
\end{equation}
where $<n>$ is the mean size of the clusters. Note that, for Metropolis, $<n>=1$ and $1$ $MCS$
is the ``time'' taken to try to flip $N$ spins, as usual.

In this work, this rescaling has been
done and all times are expressed in $MCS$. The function $\rho(t)$
is defined as:
\begin{eqnarray}
   \rho(t) & = & \int \left[ \Phi(t') - <\Phi> \right] \left[ \Phi(t'+t) - <\Phi> \right] dt' \nonumber \\ 
   & = & \int \left[ \Phi(t') \Phi(t'+t) - <\Phi>^2 \right] dt', \label{eq:autocorrelation}
\end{eqnarray}
where $\Phi(t)$ is some physical quantity.
Of course, time is a discrete quantity in the simulations; therefore, we have to 
discretize the previous equation, which
leads to \cite{newman}:
\begin{eqnarray}
   \rho(t) & = & \frac{1}{t_{max}-t} \sum_{t'=0}^{t_{max}-t} \left[ \Phi(t') \Phi(t'+t) \right] - \nonumber \\ 
    & & \frac{1}{(t_{max}-t)^2} \sum_{t'=0}^{t_{max}-t}  \Phi(t') \times \sum_{t'=0}^{t_{max}-t} \Phi(t'+t)
     \label{eq:discrete}
\end{eqnarray}

  The autocorrelation function is expected to behave, as a function of time, as
\cite{newman}
\begin{equation}
   \rho(t) = A e^{-t/\tau}, \label{eq:functionoftime}
\end{equation}
at least in its simplest form. It  is known that, in some cases, more than one
exponential term is required \cite{wansleben}; we will comment on this later.
Usually, one can measure $\tau$ from the slope of an adjusted straight line in
a semi-log plot of the autocorrelation
function versus time. However, the autocorrelation function is not well behaved
for long times, due to bad statistics (this is evident from Eq. \ref{eq:discrete},
since few ``measurements'' are available for long times). Therefore, one has to choose the region where the
straight line will be adjusted very carefully and it turns out that the
value of $\tau$ so obtained is strongly dependent on this choice.
Alternatively, one can integrate $\rho(t)$, assuming
a single exponential dependence on (past and forward) time, and obtain:
\begin{equation}
   \tau = \frac{1}{2} \int_{-\infty}^{\infty} \frac{\rho(t)}{\rho(0)} dt, \label{eq:tau}
\end{equation}
with:
\begin{equation}
   \rho(t) \equiv e^{-|t|/\tau}.
\end{equation}
   Eq. \ref{eq:tau}, when discretized, leads to \cite{salas}:
\begin{equation}
   \tau = \frac{1}{2} + \sum_{t=1}^{\infty} \frac{\rho(t)}{\rho(0)}. \label{eq:somadetau}
\end{equation}

  Of course, the sum in Eq. \ref{eq:somadetau} cannot be carried out for large values of $t$.
It has to be truncated at some point; we use a cutoff (see Ref.\cite{salas} and references
therein), defined as the value in time where the noise in the data is clearly greater 
than the signal itself. With the value of $\tau$ obtained as explained above, we made the integral of $\rho(t)/\rho(0)$
from the value of the cutoff to infinity. A criterion to accept the cutoff is that the value of this
integral is smaller than the statistical uncertainty in calculating $\tau$. Since the value
we obtain for $\tau$ is underestimated, this criterion is a safe one.

\begin{figure}
\epsfxsize=0.6\textwidth
\begin{center}
\leavevmode
\epsffile{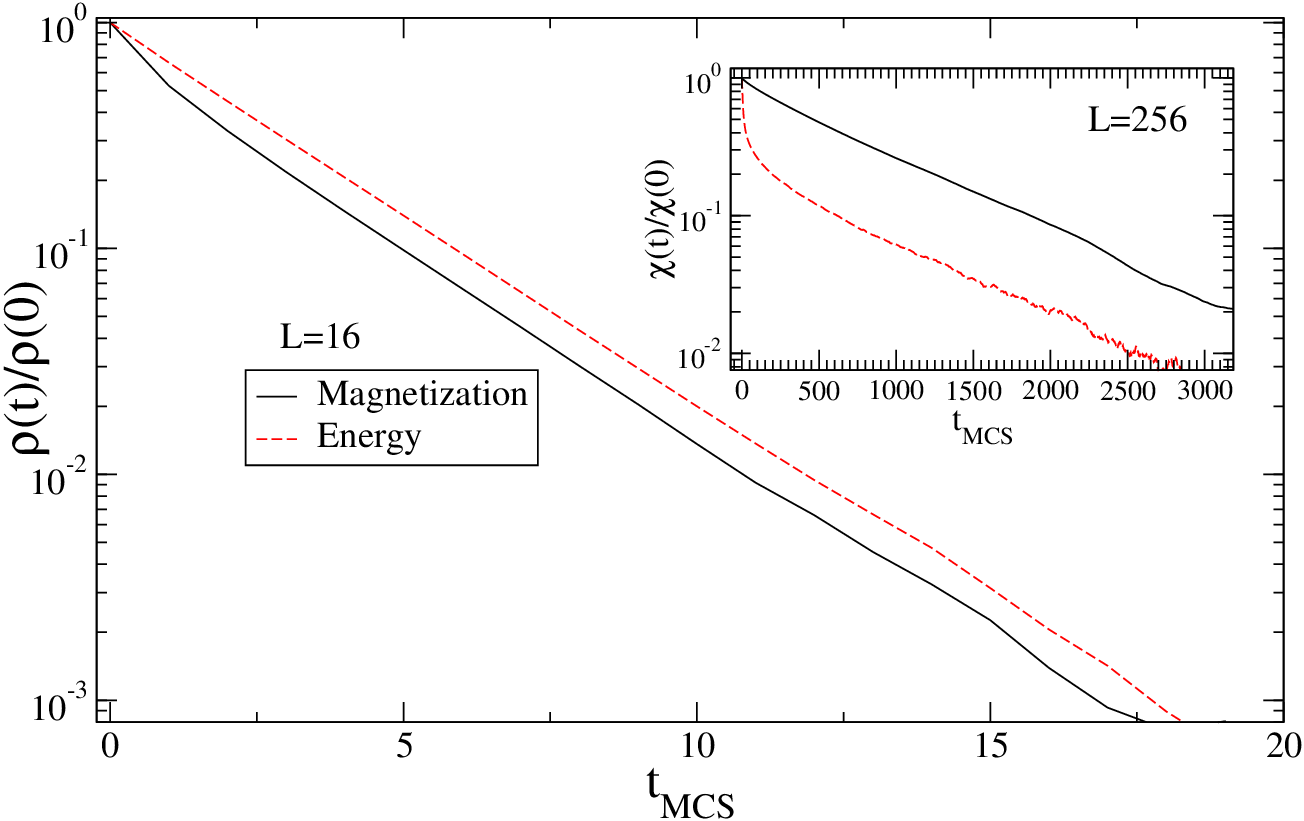}
\caption{Magnetization and energy autocorrelation functions versus time (in $MCS$)
for the Niedermayer algorithm with $E_0=0.3$ (see text). The main graph represents the behavior for
linear size $L=16$, while the inset applies to $L=256$.}
\label{fig:magnatizacaoeenergia} %fig1
\end{center}
\end{figure}

   Whenever possible, we fitted the autocorrelation time to the expected behavior, namely
$\tau \sim L^z$, in the critical region, where $z$ is the dynamic exponent. A point
worth mentioning is that the autocorrelation 
function of different quantities may behave in different ways. A typical example
is shown in Fig. \ref{fig:magnatizacaoeenergia}, where both the magnetization 
and the energy autocorrelation functions are depicted as functions of time, for the
Niedermayer
algorithm with $E_0=0.3$ and linear sizes $L=16$ (main graph) and $L=256$ (inset).
Note the abrupt drop of the magnetization autocorrelation function for small times and $L=16$. 
This is an indication that this function is not properly described
by a single exponential. On the other hand, the energy autocorrelation time
follows a straight line even for the smallest times. Therefore, we should calculate $\tau$ from the
latter, for $L=16$, using Eq. \ref{eq:somadetau}.
However, when $L$ is increased, the picture changes and now  the magnetization
autocorrelation function is well described by a single exponential (for small and intermediate values of
time), as depicted in the inset of Fig. \ref{fig:magnatizacaoeenergia}. Whenever a
crossover like this is present, we measure the dynamic exponent from the behavior 
for large values of $L$ and for the function which is well described by a single exponential for this
range of $L$, using Eq. \ref{eq:somadetau}.
But note that, for intermediate values of $t$,  
the slopes of both curves in Fig. \ref{fig:magnatizacaoeenergia} (main graph and inset) appear to be the same. However, we have already
commented on the drawback of calculating $\tau$ from the slope of the autocorrelation function on a semi-log graph.
As final notes, we would like to mention that we used helical boundary conditions and 20 independent runs (each with a
different seed for the random number generator) were made for each $E_0$ and $L$. For each seed, at least
$4 \times 10^6$ trial flips were made, in order to calculate the autocorrelation functions and their respective autocorrelation
times. The values we quote are the average of the values obtained for each seed of the random number generator and
the uncertainty in $\tau$ is the standard deviation of these $20$ values.

\section{Results and Discussion} \label{sec:results}

    \subsection{Ising model} \label{subsec:Ising}
    
    We first present our results for the Ising model and leave to the next subsection the discussion of
the results for the $XY$ model.
    
   As already discussed, the case $E_0=-1$ corresponds to the Metropolis algorithm.
 At the critical temperature, the autocorrelation time scales with $L$ as $\tau \sim L^z$, with
$z=2.1665 \pm 0.0012$ \cite{nightingale}. We have simulated this case only as a test for our
algorithm. The value we found for $z$ is consistent with the one quoted above and the
scaling law is obeyed, even for the smallest values of $L$ we simulated. Note also that,
for the Metropolis algorithm ($E_0=-1$), it is the magnetization autocorrelation time which is well described 
by a single exponential.

     The first non-trivial value of $E_0$ we simulated was $-0.9$. In Fig. \ref{fig:tao-0,9} the autocorrelation
 times for the magnetization is depicted as function of $L$. We note that, for this value of $E_0$,
 only the autocorrelation function for the magnetization is well described by a single exponential. The
 initial decay of the corresponding function for the energy has an abrupt drop for small times.Therefore,
it is not a reliable quantity to extract the autocorrelation time from. The value of $z$ was obtained from the
curve for the magnetization and its value is $z=2.16 \pm 0.04$, which is, within error bars, the same value
as for the Metropolis algorithm. 
 
\begin{figure}
\epsfxsize=0.6\textwidth
\begin{center}
\leavevmode
\epsffile{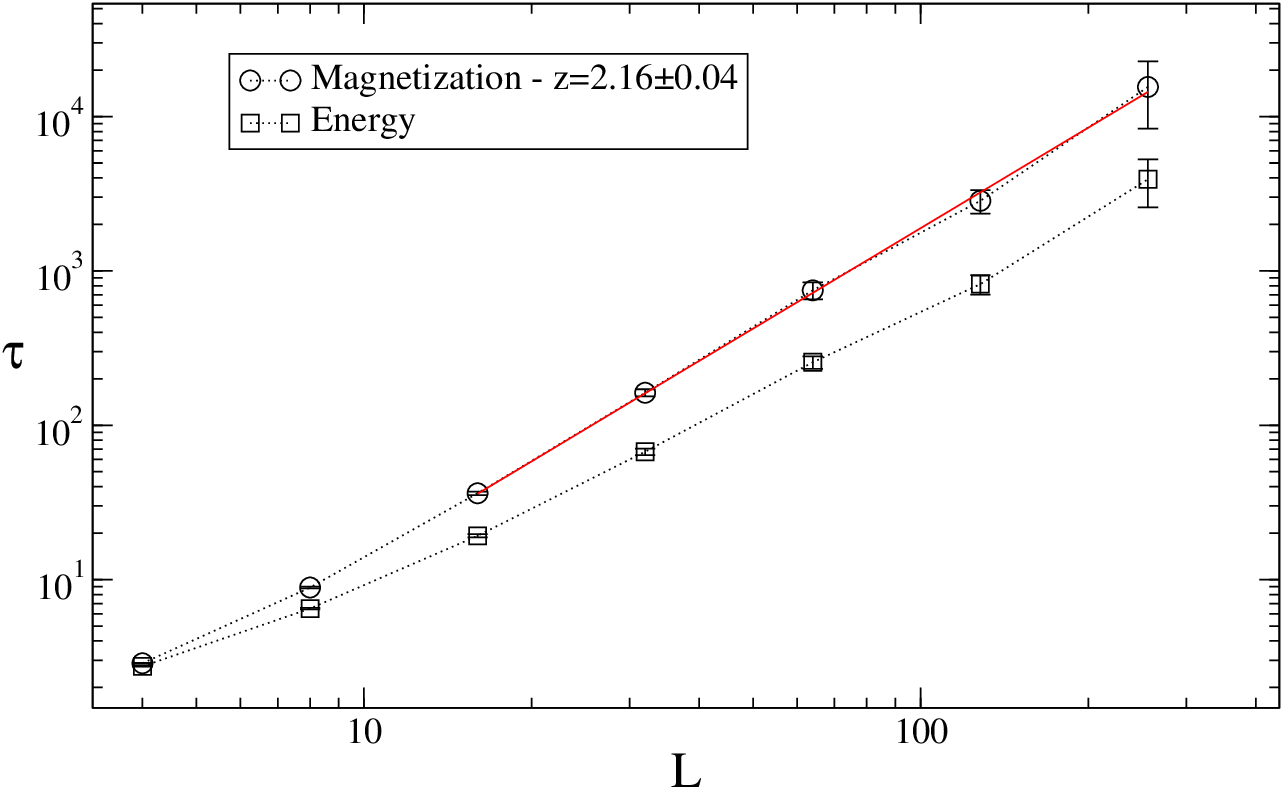}
\caption{Log-log graphs of magnetization (\opencircle) and energy (\opensquare) autocorrelation time
(in $MCS$) versus linear size $L$
for the Niedermayer algorithm with $E_0=-0.9$. The quoted value for $z$ is obtained from the slope
of an adjusted straight line for the magnetization autocorrelation time for $L \ge 16$ (see text). The dotted line
is just a guide to the eye.}
\label{fig:tao-0,9} %fig2
\end{center}
\end{figure}

    In Fig. \ref{fig:nmedio-0,9} the behavior of the mean size of the clusters of spins, $<n>$, is shown, as function of
$L$.  For this value of $E_0$, it seems that $<n>$ does not change with $L$. We will see shortly that in fact it
initially grows with $L$ and eventually saturates at some value of $L$, which we call $\widetilde{L}$.
    
 \begin{figure}
\epsfxsize=0.6\textwidth
\begin{center}
\leavevmode
\epsffile{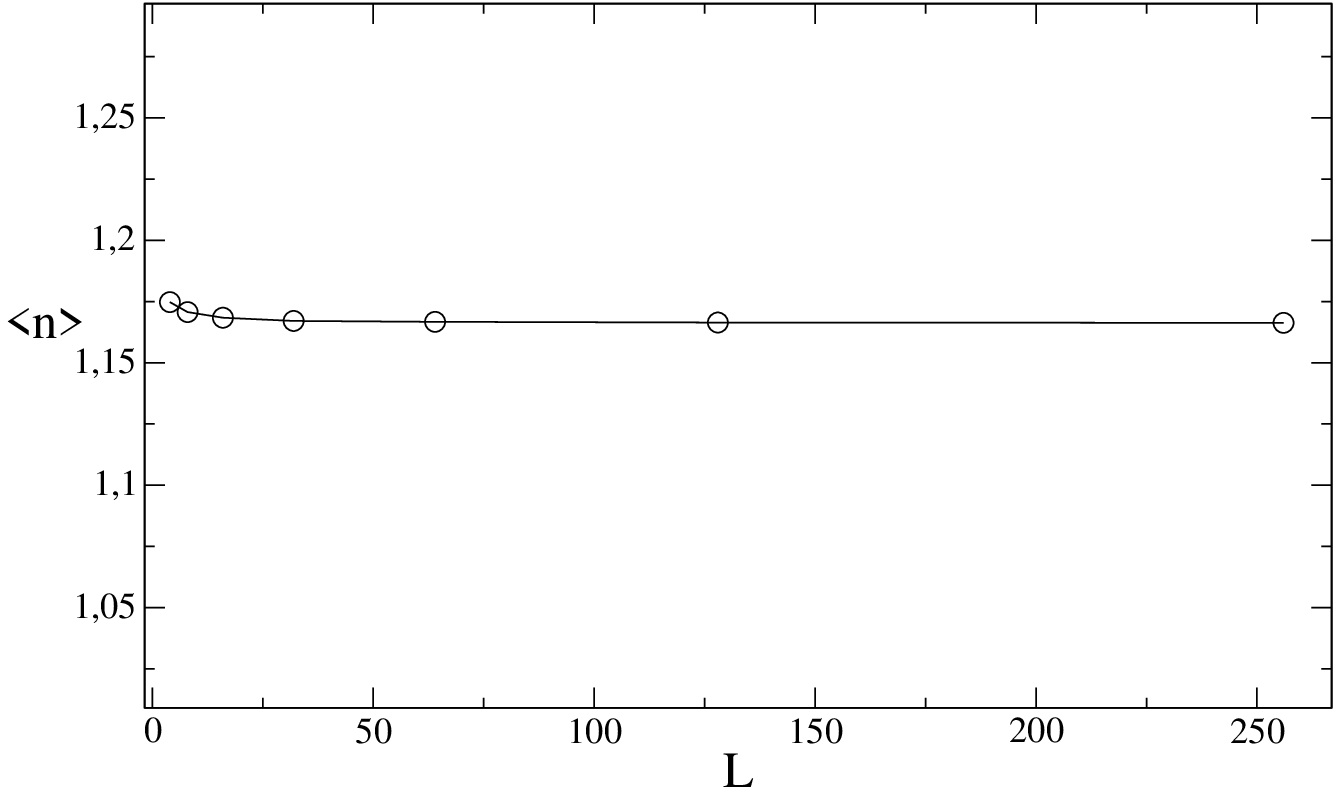}
\caption{Mean size of the clusters of possibly flipped spins as function of the linear size $L$ for $E_0=-0.9$.}
\label{fig:nmedio-0,9} %fig3
\end{center}
\end{figure}

The overall picture does not change for $E_0=-0.5$: the magnetization autocorrelation function is well described by a single exponential law and the autocorrelation time was calculated from it. The dynamic exponent is $z=2.12 \pm 0.03$, still
consistent with the Metropolis value (the error bars we quote are all one standard deviation; the intersection with the 
expected value for the Metropolis algorithm, for this case, is obtained assuming two standard deviations for the error). 
Since the picture for $E_0=-0.5$ does not change
from the one for $E_0=-0.9$, we will not depict the graphs for the former.

   For $E_0=0$, a crossover clearly takes place, as shown in Fig. \ref{fig:tao0,0}: for small $L$,
the energy autocorrelation times are larger than their magnetization counterparts, while the situation
is reversed for larger $L$ (this behavior is more evident for $E_0=0.3$; we showed the corresponding graph
in Fig. \ref{fig:magnatizacaoeenergia} above and will comment on it below). The value of $z$ is obtained from the slope
of an adjusted straight line for the magnetization autocorrelation function, for values of $L$ beyond the point where
the crossover takes place. It reads $z=2.15 \pm 0.01$ in this case, again compatible with the Metropolis value.
The behavior of $<n>$ is shown in Fig. \ref{fig:nmedio0,0}: 
it grows initially with $L$ but eventually saturates
at $\widetilde{L} \sim 15$. For small values of $L$ it is the autocorrelation function for the energy
which is well described by a single exponential, while the corresponding function for the magnetization shows an
abrupt drop for small times. The situation is reversed for $L>\widetilde{L}$.

\begin{figure}
\epsfxsize=0.6\textwidth
\begin{center}
\leavevmode
\epsffile{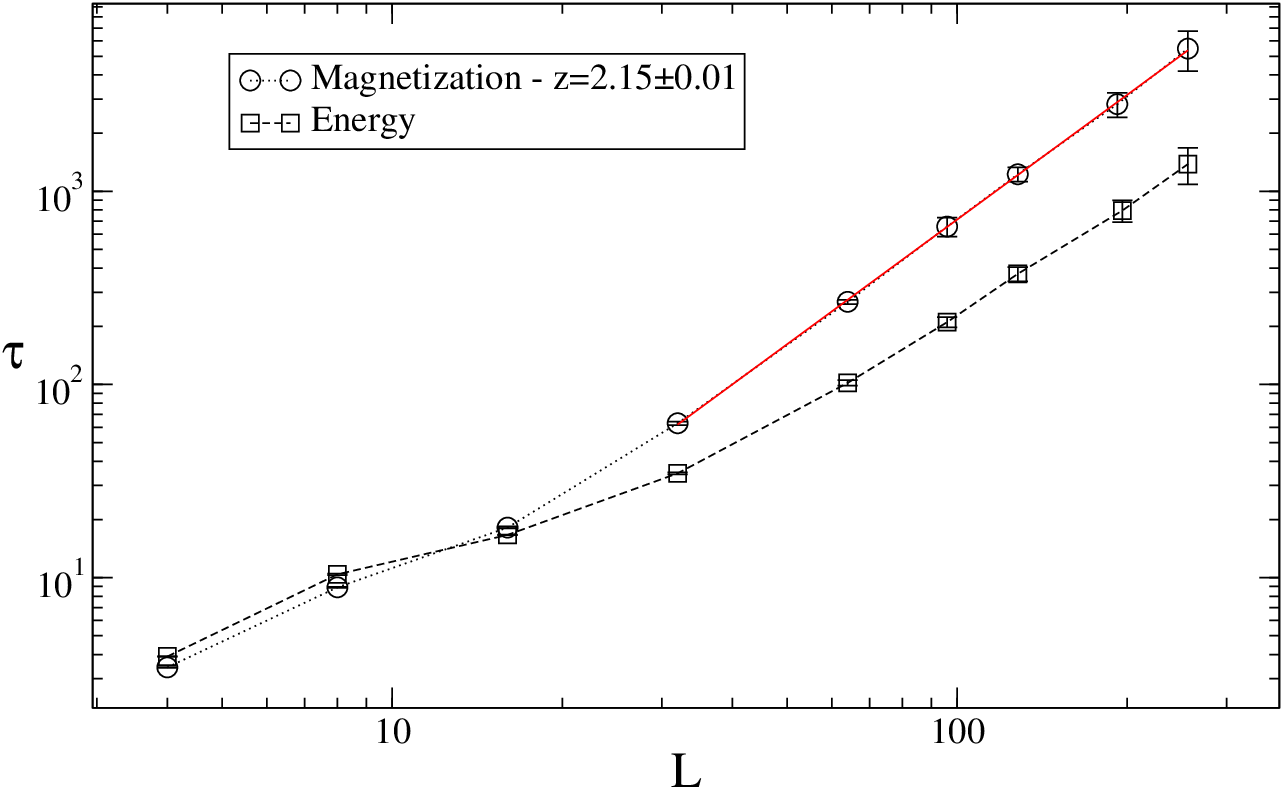}
\caption{Log-log graphs of magnetization (\opencircle) and energy (\opensquare) autocorrelation time
(in $MCS$) versus linear size $L$
for the Niedermayer algorithm with $E_0=0.0$. The quoted value for $z$ is obtained from
the slope
of an adjusted straight line for the magnetization autocorrelation time, for values of $L$
beyond the point where
the crossover takes place. The dotted line is just a guide to the eye.}
\label{fig:tao0,0} %fig4
\end{center}
\end{figure}

\begin{figure}
\epsfxsize=0.6\textwidth
\begin{center}
\leavevmode
\epsffile{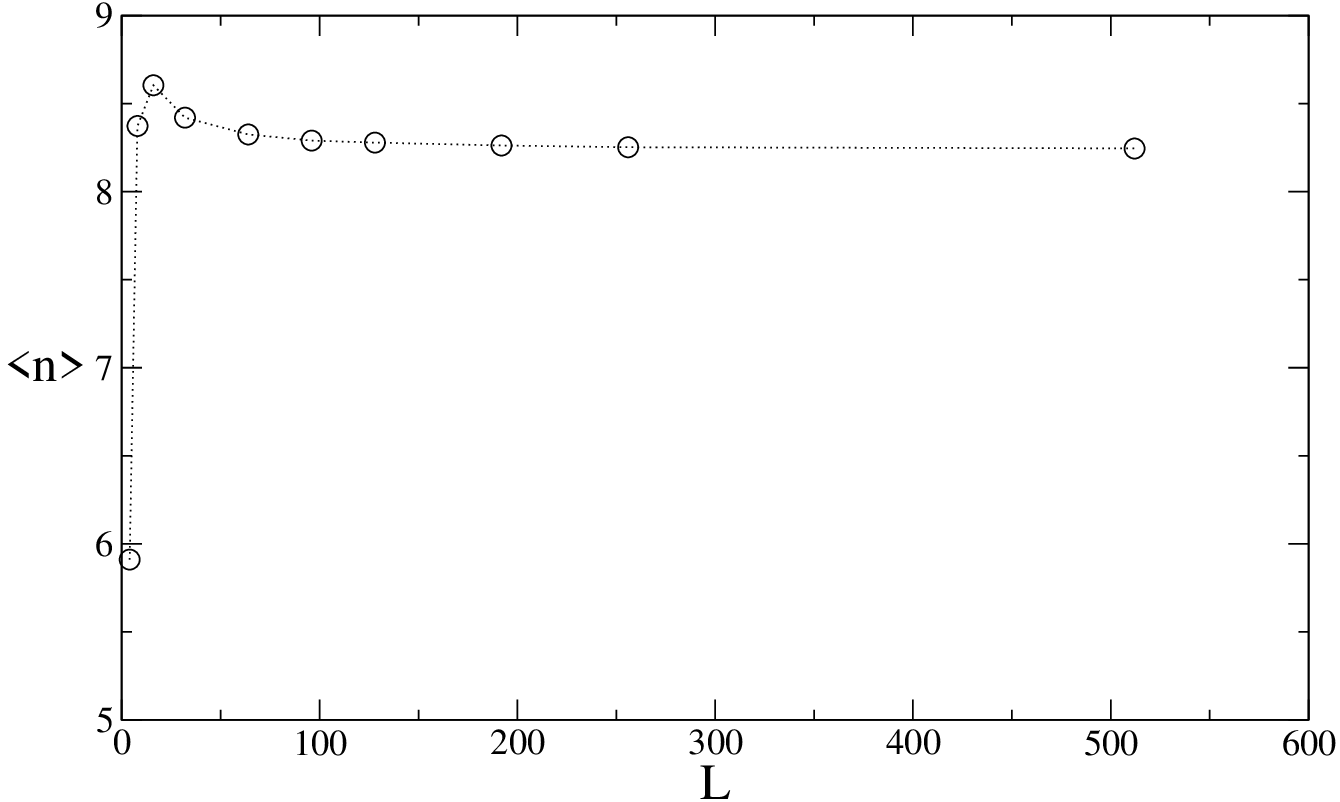}
\caption{Mean size of the clusters of possibly flipped spins as function of the linear size $L$ for $E_0=0.0$.}
\label{fig:nmedio0,0} %fig5
\end{center}
\end{figure}

      This picture is maintained for $E_0 > 0.0$, with the value of $\widetilde{L}$ increasing with $E_0$ and the
crossover taking place at larger and larger values of $L$.
The dynamic exponent $z$ is given by $2.16 \pm 0.03$ and $2.12 \pm 0.04$ for $E_0=0.3$ and $0.5$,
respectively. Both are compatible with the value for the Metropolis algorithm. 
   
     In Fig. \ref{fig:magnatizacaoeenergia} we show the change in the behavior of the autocorrelation
functions for the magnetization and the energy. Note that the crossover mentioned above is connected also to
the possibility of describing the autocorrelation function by a single exponential: this is accomplished by the
energy autocorrelation function for small values of $L$ and for its magnetization counterpart for larger values
of $L$. 
     
     Finally, for $E_0=0.7$ and $0.9$ the 
crossover happens at values of $L$ large enough to prevent a reliable estimate of $z$. It is necessary to go
to values of $L$ well above our present computational capabilities to be able to extract $z$ from the graphs.

     Nevertheless, the overall trend is well determined: for $0 \le E_0 < 1$, the dynamic behavior is the Metropolis'
one but this behavior sets in only for large enough $L$. The size of the clusters of turned spins increases with
$E_0$ but eventually saturates for $L=\widetilde{L}$, where $\widetilde{L}$ increases with $E_0$.
For $L>\widetilde{L}$, the relative size of the clusters (i.e, the ratio $<n>/L^2$) decreases and, in this sense, the algorithm is like
a single-spin one (Metropolis, in our case), explaining the value of its dynamic exponent. Therefore, the Wolff algorithm
(corresponding to $E_0=1$) is still the best choice, when compared to the Niedermayer algorithm with $E_0<1$.

     We postpone the discussion of the Wolff algorithm and go to $E_0>1$ . In this
 case, spins in different states may be part of the same cluster, although with a smaller probability than spins in the same
 state, and a cluster will always be flipped (see $(iii)$ on page $2$). For $E_0 \gg 1$, almost all spins in the finite lattice will take part in the
 cluster and the algorithm will not be optimal (in fact, it won't even be ergodic for $E_0 \rightarrow \infty$). Therefore, if
 the Niedermayer algorithm is more efficient than Wolff's, it should be for $E_0$ close to $1$. We, therefore, studied the
 cases $E_0=1.1$ and $1.05$. The results are qualitatively equivalent and in Fig. \ref{fig:tao-1,1-1,05} we show both.
 Note that the growth of $\tau$ with $L$ is faster than a power law for both values of $E_0$. In the inset, we show the corresponding
 graph for $E_0=1.1$: a crossover is also present but the value where it takes place decreases with $E_0$ and
 for $E_0=1.1$ it is not seen. Since the value of the autocorrelation time is already greater then for the Wolff algorithm,
 for a given $L$, and it grows faster than a power law with $L$, again the optimal algorithm is Wolff's.
 
 \begin{figure}
\epsfxsize=0.6\textwidth
\begin{center}
\leavevmode
\epsffile{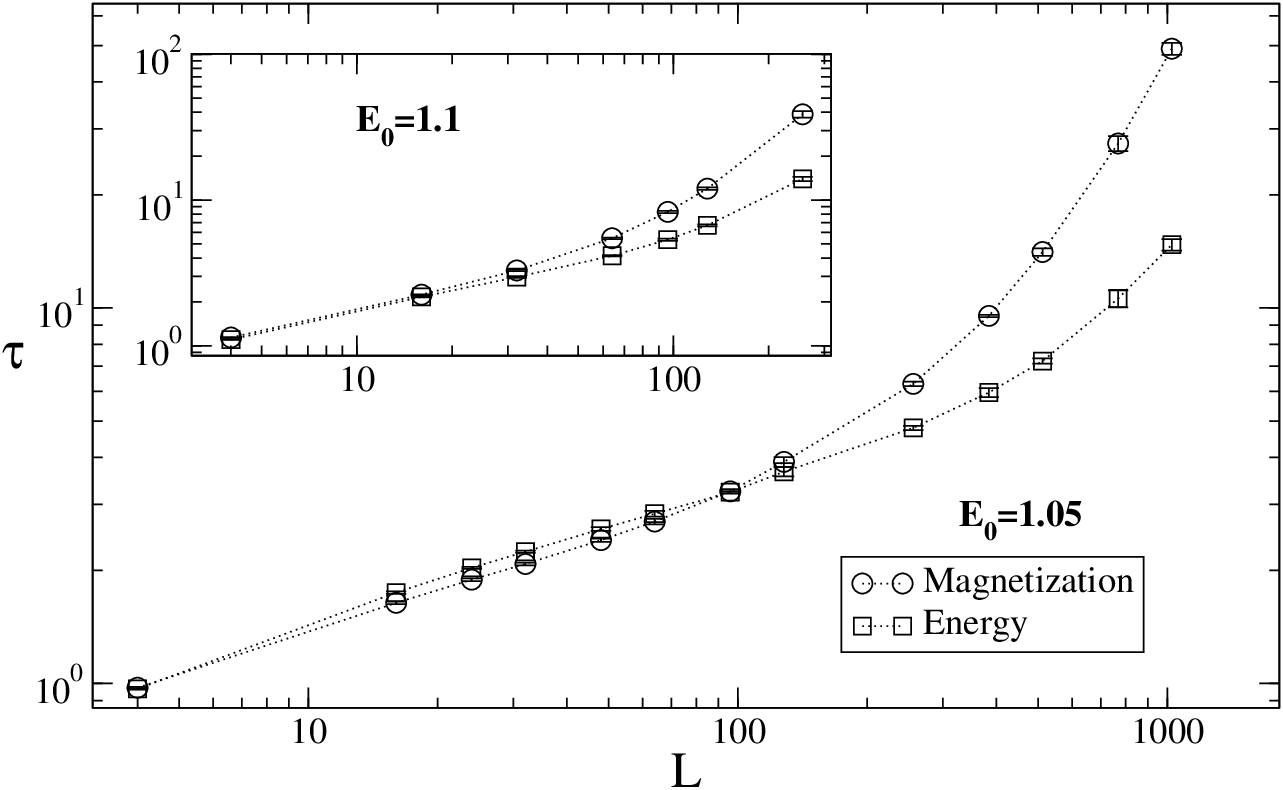}
\caption{Log-log graphs of magnetization and energy autocorrelation time (in $MCS$)
versus linear size $L$
for the Niedermayer algorithm with $E_0=1.05$ and $E_0=1.1$ (inset). In both graphs the
autocorrelation time $\tau$ is plotted
as function of the linear size $L$. The dotted line is just a guide to the eye.}
\label{fig:tao-1,1-1,05} %fig6
\end{center}
\end{figure}

\begin{figure}
\epsfxsize=0.6\textwidth
\begin{center}
\leavevmode
\epsffile{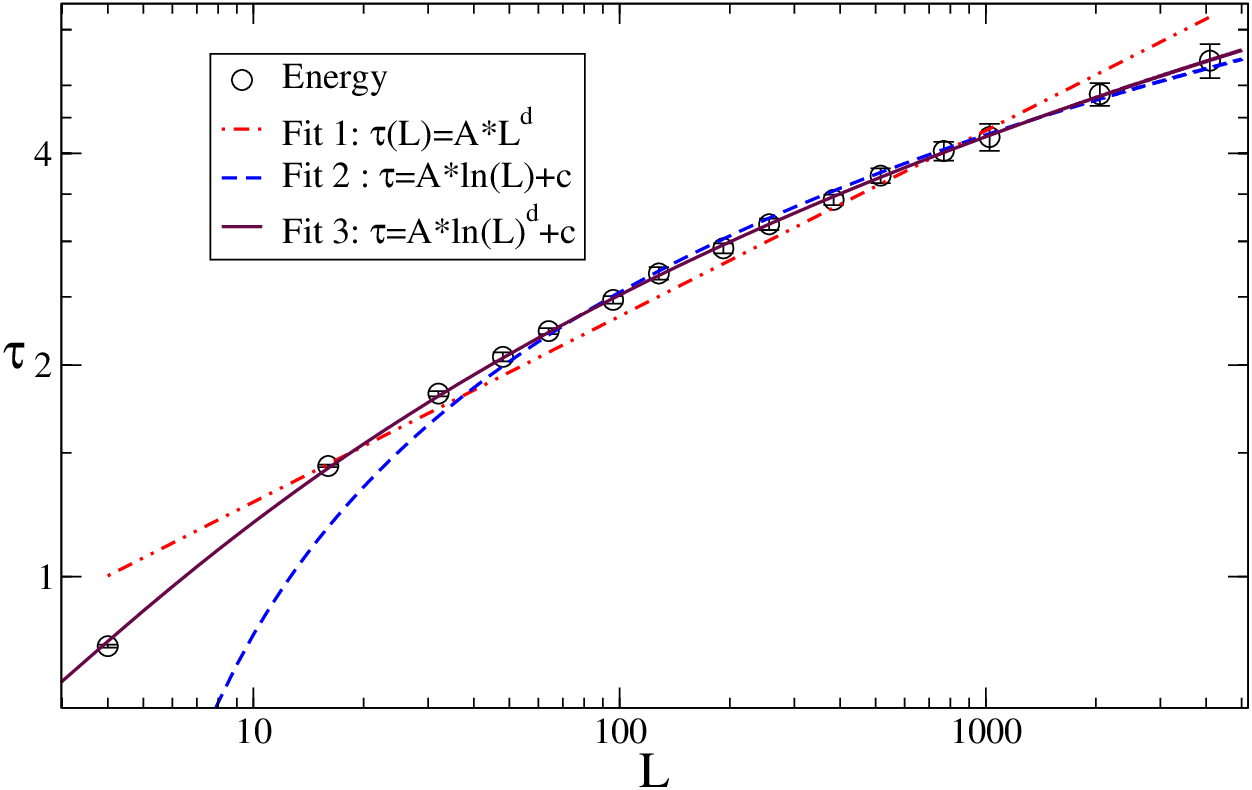}
\caption{Log-log graph of the energy autocorrelation time (in $MCS$) as
a function of $L$
for the Wolff algorithm. The three fitted curves proposed in \ref{eq:ajustes} are showed.}
\label{fig:wolff_final} %fig7
\end{center}
\end{figure}

     We single out the discussion of the Wolff algorithm ($E_0=1$), in order to compare with other values
 of $E_0$. Although our intention is not to calculate a precise value of $z$ for this algorithm, we have adjusted the autocorrelation
 time for the energy (in this case, it is this function which is well described by a single
 exponential for small values of the time) as a function of $L$ for three different functions, namely:
\begin{equation}
    \tau =    \left\{   \begin{array}{l}
                                      A  L^z \\
                                      A  \ln L + C \\
                                      A (\ln L)^z+C
                           \end{array}    \right.  
     \label{eq:ajustes}
 \end{equation}
  The first function is the usual scaling law assumed for the autocorrelation function at
the critical point. We can see in Fig. \ref{fig:wolff_final} that there is no indication that
the best adjusted curve will eventually be a straight line, in a log-log plot. Since previous
calculations tend to point to a value of $z$ close to zero
for the Wolff algorithm in two dimensions, one cannot
exclude the possibility
of a logarithmic dependence, which is the case of the second function in the above
equation. The fitting is better than for the power law but it is not a satisfactory one either. 
Moreover, it tends to deviate from the data for large enough $L$. The third
function is an {\it ad hoc} assumption, which proved to be the best fit to our data, as can be
seen in Fig. \ref{fig:wolff_final}. The parameters of the function are obtained from a non-linear
fitting:
\begin{equation}
     \tau=A (\ln L)^z+C,
     \label{eq:best_fit}
\end{equation}
with $A=0.21\pm0.01$, $z=1.50\pm0.02$ e $C=0.47\pm0.03$. We have no theoretical explanation 
for this behavior. The constant $C$, however, is a finite-size correction. The
behavior in Eq. \ref{eq:best_fit} is expected to hold true for {\it large} enough $L$ and the logarithmic dependence
makes the scaling region to be reached only for very large values of $L$. For this region, one would
expect a simpler law, namely $\tau = A (\ln L)^z$; however, for intermediate or small values of $L$, the
constant $C$ acts as a finite-size correction. A similar scaling law was found for the exponential relaxation
time for the Swendsen-Wang algorithm \cite{du}.
  
\begin{figure}
\epsfxsize=0.6\textwidth
\begin{center}
\leavevmode
\epsffile{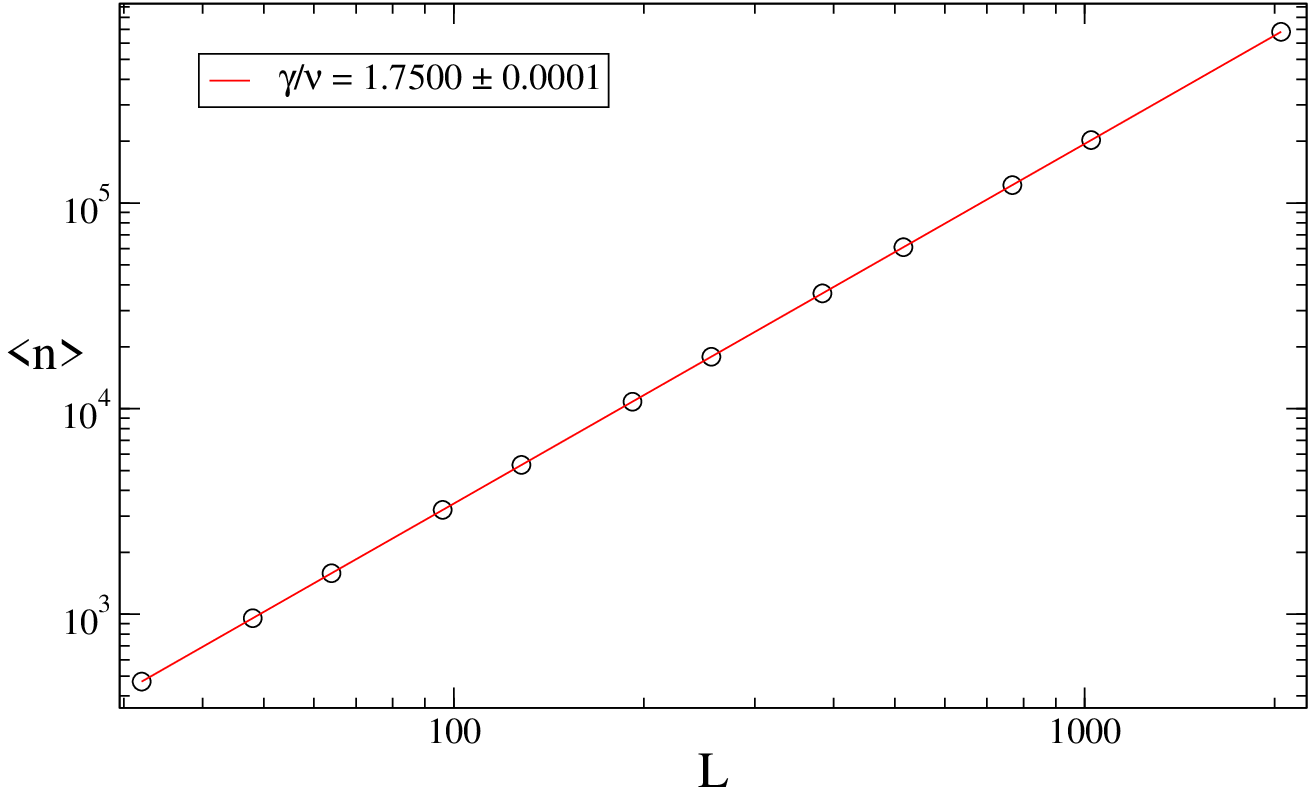}
\caption{Log-log graph of the mean size of turned clusters as function of $L$. The slope
is an evaluation of $\gamma/\nu$. }
\label{fig:nmedio_Wolff} %fig8
\end{center}
\end{figure}

     We also depict the mean size of the clusters of turned spins, $<n>$, as a function of $L$ in Fig. \ref{fig:nmedio_Wolff}. The
slope of the straight line is $1.7500 \pm 0.0001$, which is, as expected \cite{newman}, the value
for the ratio $\gamma/\nu$. Note that, contrarily to what happens for $E_0 < 1$, there is
no saturation
of $<n>$ with $L$. This seems to explain why Wolff and Niedermayer algorithms are in different dynamic universality
classes.
    
   \subsection{XY model} \label{subsec:XY}
   
    We have applied the Niedermayer algorithm in the study of the dynamic behavior of the $XY$ model as well.
The generalization of this algorithm to continuous models is outlined in the Appendix. 
Although we have studied three values of $E_0$, our results are conclusive and lead to an overall picture 
which is analogous to the one for the Ising model.
We have used the value $k_B T_c/J= 0.8865$ for the transition temperature of the two-dimensional $XY$ model. 
This value is only $0.7 \% $ off of the most recent evaluation of $k_B T_c/J$ for this model \cite{arisue}. 

\begin{figure}
\epsfxsize=0.6\textwidth
\begin{center}
\leavevmode
\epsffile{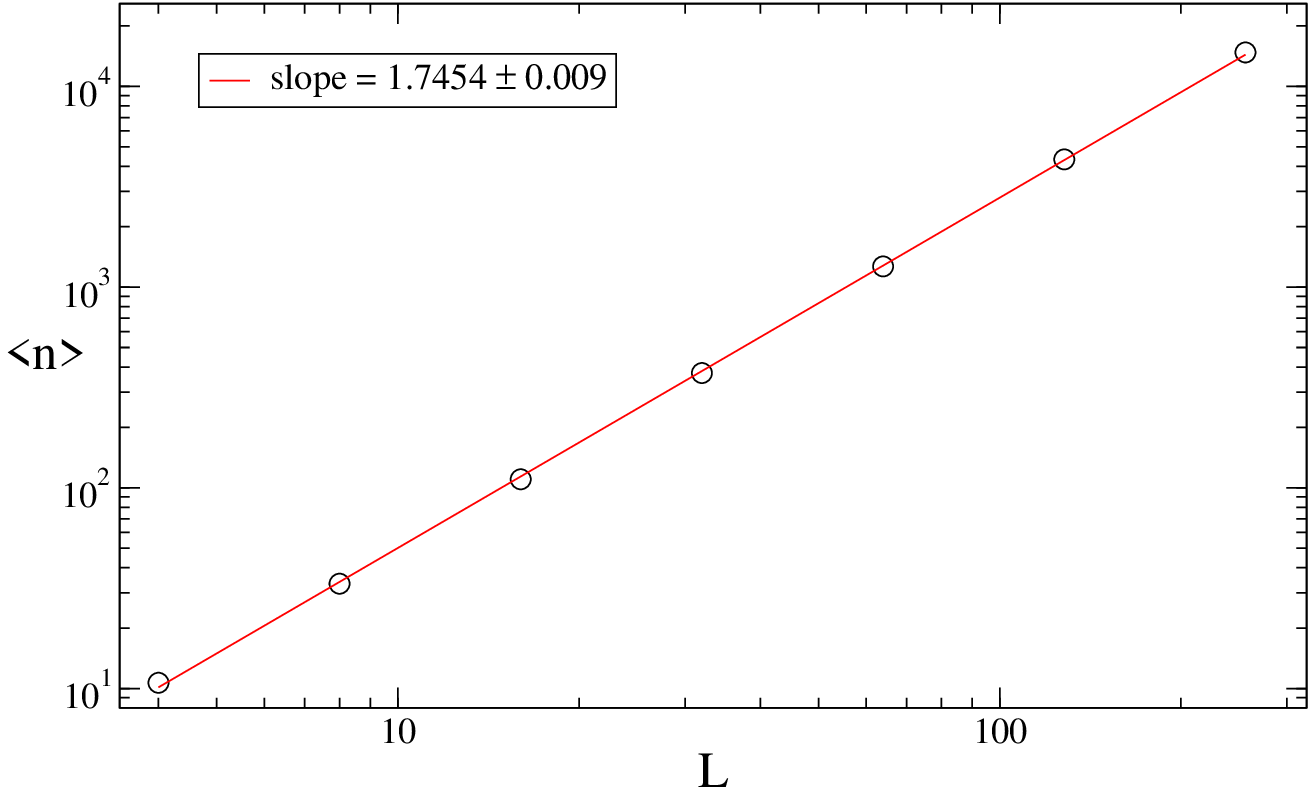}
\caption{Log-log plot of the mean size of the clusters of flipped spins for the $XY$ model
versus the linear size of the lattice. The slope of the curve just misses the expected
value for  $2-\eta$, $7/4$ \cite{butera}.}
\label{fig:nmedio-XY-E1.0} %fig9
\end{center}
\end{figure}

   The mean size of the clusters of flipped spins for the Wolff algorithm as a function of the linear size of the
lattice is depicted in Fig. \ref{fig:nmedio-XY-E1.0}. The slope of the straight line is
$1.7454 \pm 0.009$, which is slightly different from the expected value for $2-\eta$ for this model
at $k_B T_c/J$, $7/4$ \cite{butera}. In fact, the small discrepancy may be due to the fact that we are not
using the (unknown) exact value for the transition temperature. 
   
\begin{figure}
\epsfxsize=0.6\textwidth
\begin{center}
\leavevmode
\epsffile{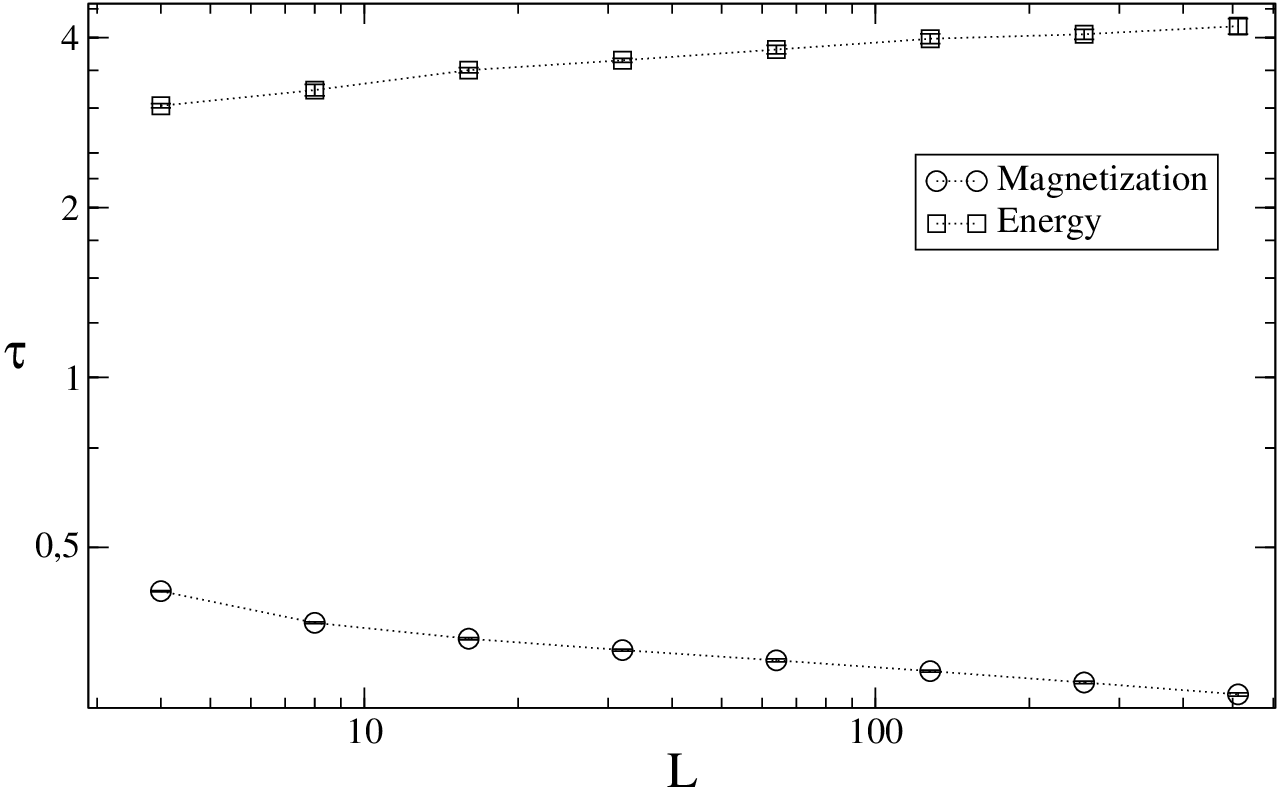}
\caption{Autocorrelation times for the magnetization (circle) and energy (square)  for the
Wolff algorithm
applied to the two-dimensional $XY$ model.} 
\label{fig:tao-XY-E1.0} %fig10
\end{center}
\end{figure}

   The autocorrelation times for the magnetization and energy for the Wolff algorithm 
 are shown in Fig. \ref{fig:tao-XY-E1.0}.
We have not tried to fit the data but it is evident that the
energy autocorrelation time grows with $L$ slower than a power law. The \textit{decrease} in the magnetization autocorrelation
time has been observed previously (in fact, an oscillation was observed in an algorithm which
mixed Wolff's and Swendsen-Wang's procedures but the overall picture is qualitatively similar
to ours; see Ref.\cite{edwards}).

\begin{figure}
\epsfxsize=0.6\textwidth
\begin{center}
\leavevmode
\epsffile{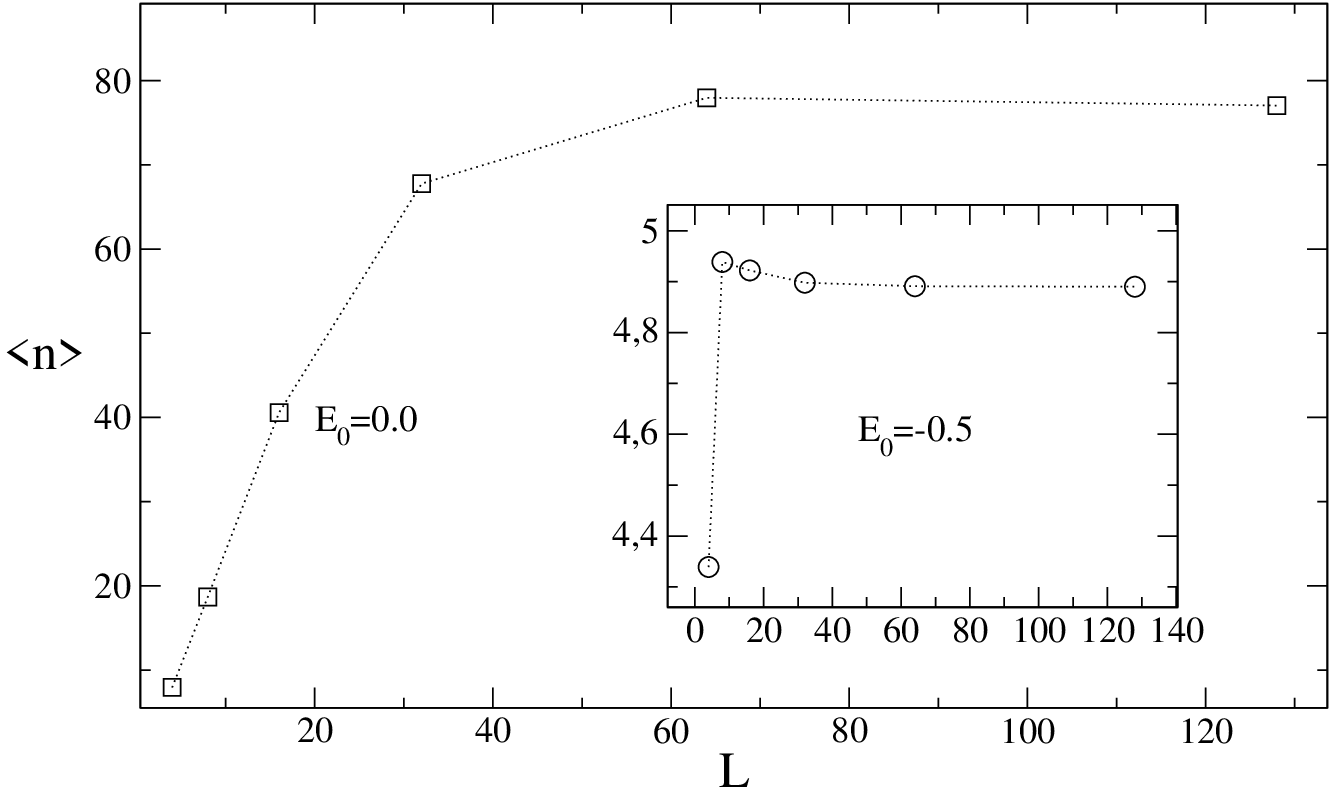}
\caption{Mean size of the clusters of flipped spins for $E_0=-0.5$ (inset) 
and $E_0=0$ (main graph)
for the two-dimensional $XY$ model.} 
\label{fig:nmedio-XY-0.0_-0.5} %fig11
\end{center}
\end{figure}

\begin{figure}
\epsfxsize=0.6\textwidth
\begin{center}
\leavevmode
\epsffile{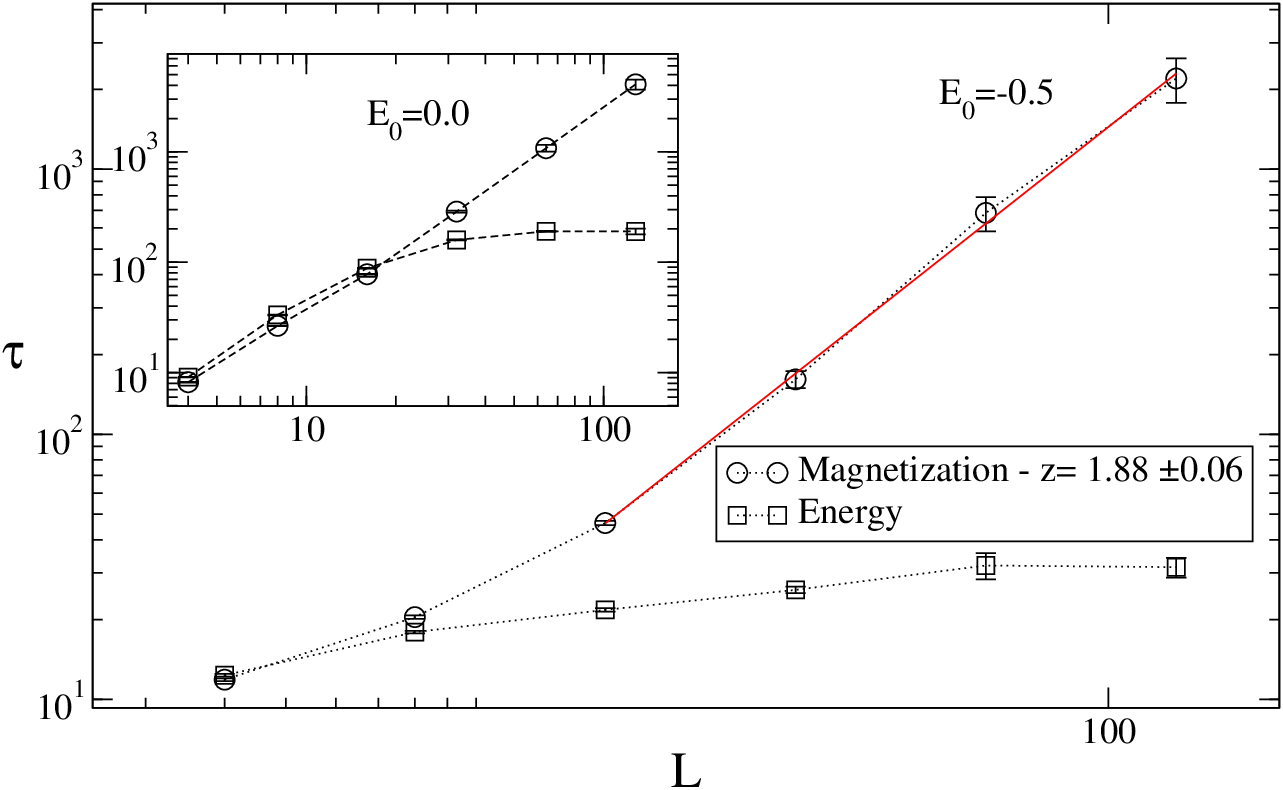}
\caption{Autocorrelation time for the magnetization and energy as a function of $L$ for $E_0=-0.5$ (main graph) 
and $E_0=0$ (inset)
for the two-dimensional $XY$ model.} 
\label{fig:tao-XY-0.0_-0.5} %fig12
\end{center}
\end{figure}

\begin{figure}
\epsfxsize=0.6\textwidth
\begin{center}
\leavevmode
\epsffile{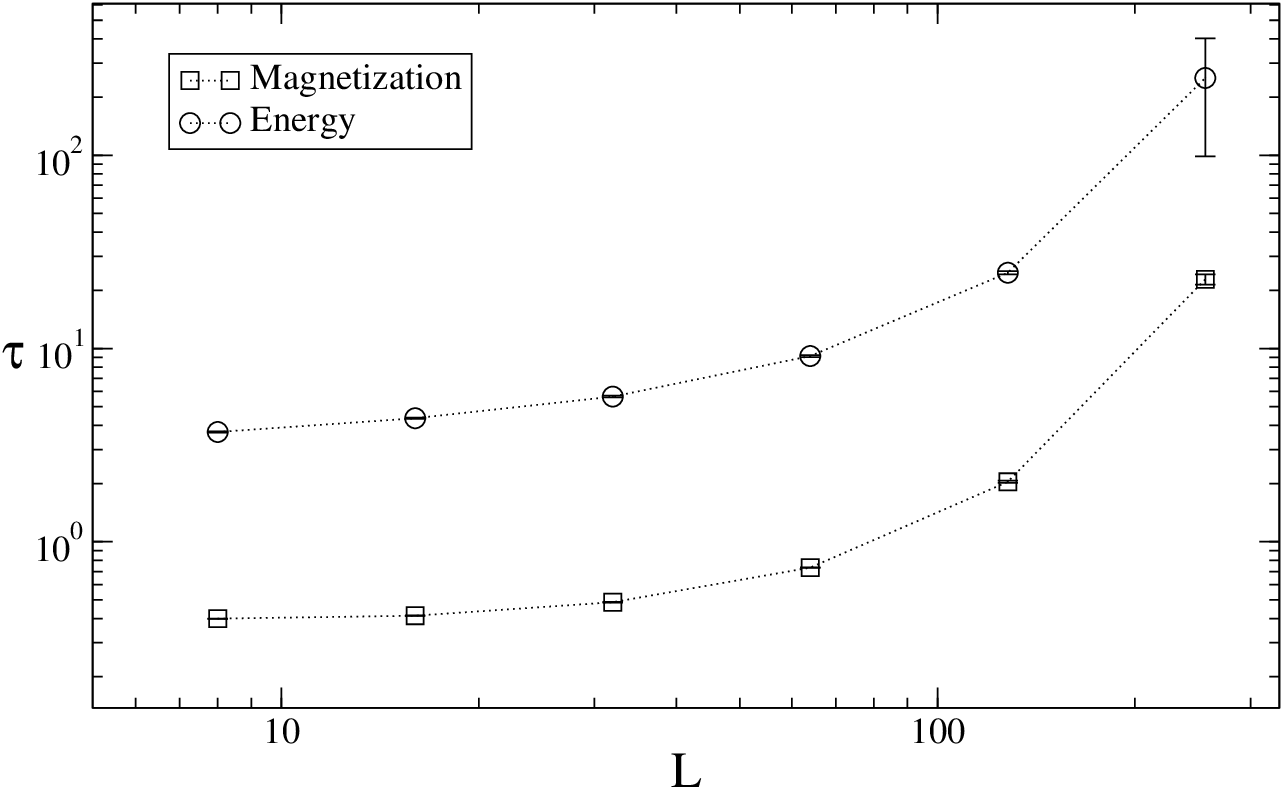}
\caption{Autocorrelation time for the magnetization and energy as a function of $L$ for $E_0=1.05$
for the two-dimensional $XY$ model.} 
\label{fig:tau-XY-1.05} %fig13
\end{center}
\end{figure}

    We have simulated also the cases $E_0=0$ and $E_0=-0.5$.  The mean size of clusters
of  flipped spins saturates and the value of saturation increases with $E_0$ (see
Fig. \ref{fig:nmedio-XY-0.0_-0.5}). Therefore, one expects the same picture as for the Ising model: 
in particular, the dynamic behavior
for $L$ large enough is the Metropolis' one. This is confirmed for $E_0=0.0$ explicitly,
where the dynamic exponent measured is $z=1.916 \pm 0.004$ (see Fig. \ref{fig:tao-XY-0.0_-0.5}). Recalling our reasoning for the Ising model for $E_0<1$,
we can infer that the value just quoted for $z$ is an evaluation of the dynamic exponent for the Metropolis
algorithm applied to the two-dimensional $XY$ model. Since, to the best of our knowledge, there is no previous evaluation
of $z$ for this model and for the Metropolis algorithm, we have made a crude
evaluation of $z$ for this case and obtained the value $1.89 \pm 0.03$, which is in agreement, within error bars,
with the value we obtained for $E_0=0.0$ for large $L$.
Clearly, the Wolff algorithm is the most efficient, when compared to
the Niedermayer algorithm with the two values of $E_0$ quoted above. We have also simulated one
example of $E_0>1$, namely $E_0=1.05$. The behavior is qualitatively the same as for the Ising model; see Fig. \ref{fig:tau-XY-1.05}.
Note that the growth of the autocorrelation time is faster than a single power law; in fact, it is well fitted by an exponential.
Therefore, also for the $XY$ model the best
choice is $E_0=1$ (Wolff algorithm), when compared to the Niedermayer algorithm.
  
\section{Summary} \label{sec:summary}

   We have studied the dynamic behavior of the Niedermayer algorithm applied to the two-dimensional
Ising and $XY$ models. Our main goal is to compare its efficiency with the Wolff algorithm's. The latter is a
particular case of the Niedermayer algorithm, such that a parameter governing the size of the flipped clusters,
$E_0$, assumes the value $1$. 

   We show that, for $-1 < E_0 < 1$, the dynamic behavior eventually recovers
the Metropolis' ($E_0=-1$) one. This behavior is linked to the saturation of the mean size of the clusters, 
which happens for all $E_0 < 1$, leading to a {\it decrease} of the relative size of these
clusters when $L$ increases. 

    For the Wolff algorithm and the Ising model, we propose an scaling function for the autocorrelation time for the magnetization.
This choice is an {\it ad hoc} one but fits the data very well and does not coincide with any function
proposed so far in the literature. We were not able to make a fitting with the same statistical quality for the $XY$ model.

    For $E_0>1$, the values of the autocorrelation times are greater than those for the Wolff algorithm and
 grow faster than a power law with $L$.
    
     Therefore, at least for these two models, the Wolff algorithm is superior to Niedermayer's.
       
\section{Appendix}

The Hamiltonian for the XY model can be written as: 
\begin{equation}
 \mathcal{H}=-J\sum_{<i,j>} \vec{s_i}\cdot\vec{s_j}
     \label{eq:Hxy}
\end{equation}
where $J$ is the coupling constant and $\vec{s_i}$ is the spin of site $i$,
represented by a unit vector in any direction in the $xy$ plane. 

To start the cluster we randomly choose a preferred direction $\hat{n}$ and a
spin $\vec{s_i}$. This spin is the first one of the cluster. Neighbours of $\vec{s_i}$ are added to cluster with probability:
\begin{equation}
 P_{add} = \left\{ \begin{array}{ll}
	                     1-e^{K  E_{ij}\left(1+E_0 \right) },  & \mbox{if} \;\; E_{ij}<E_0, \\
                             0,                                      & \mbox{otherwise},
	                  \end{array}
                 \right.  \label{eq:Padd_xy}
\end{equation}
where in the XY model, $ E_{ij}=-(\vec{s_i}\cdot\hat{n})(\vec{s_j}\cdot\hat{n})$. Note that,
if $E_0 \le 1$, only sites that have the same component
in the direction $\hat{n}$ as $ \vec{s_i}$ may be
added to the cluster. The procedure for the construction of the cluster continues as
described for the Ising model. After the cluster is built the new directions of the spins are
given by a reflection with respect to axis perpendicular to $\hat{n}$. 

The acceptance ratio for XY model is slightly different from the one for the Ising model. We cannot
define the energy difference ($\Delta E$) as the number of parallel and anti-parallel
spins to the cluster. So we must calculate the energy before and after the cluster is
flipped. In this case we define de acceptance ratio, for $E_0 \le 1$, as:
\begin{equation}
 \label{aceite_xy}A(a \rightarrow b)={\Bigg \lbrace}
\begin{array}{cc}
e^{-\frac{\Delta E}{2}K(1-E_0)}, & \mbox{if} \;\;\;\Delta E > 0 \\
        1 \;\;\;\;\;\;\;\;\;\;\;\;\;\;\;\;\;\;\;\;\;, & \mbox{if} \;\;\;\Delta E < 0,
 \end{array}
\end{equation}
where $a$ and $b$ have the same  meaning as before and $\Delta E$ is the difference in
energy between configurations $a$ and $b$, in units of $J$.
As we can see, for $E_0=-1$ we regain the Metropolis algorithm with $A=e^{- K \Delta E}$ and
for $E_0=1$ we regain the Wolff algorithm with $A=1$ for all clusters. These choices ensure that detailed
balance is obeyed.

   The generalization for $E_0>1$ is analogous to the one described above and, again, spins with different signs 
for the component along $\hat{n}$ may also be part of the cluster and $A=1$ always.

\ack
The authors would like to thank the Brazilian agencies
FAPESC, CNPq, and CAPES for partial financial support.
%\end{acknowledgments}

\section*{References}
\bibliographystyle{unsrt}
\bibliography{references}

\end{document}